\documentclass[aps,prb,groupedaddress,showpacs,amsmath,amssymb,twocolumn,floatfix,superscriptaddress,10pt]{revtex4-1}
\usepackage{graphicx}
\usepackage{bm}
\usepackage{color}
\usepackage[hidelinks]{hyperref}
\hypersetup{
    colorlinks,
    citecolor=blue,
    filecolor=blue,
    linkcolor=blue,
    urlcolor=blue
}

\newcommand{\ud}{\,\mathrm{d}}
\newcommand{\im}{\textrm{i}}

\DeclareMathOperator{\e}{e}
\DeclareMathOperator{\sgn}{sgn}
\DeclareMathOperator{\Det}{Det}
\DeclareMathOperator{\Tr}{Tr}
\DeclareMathOperator{\Or}{O}

\begin{document}

\title{On the bulk boundary correspondence and the existence of Majorana bound states on the edges of 2D topological superconductors}

\author{Nicholas Sedlmayr}
\email{ndsedlmayr@gmail.com}
\affiliation{Department of Physics and Medical Engineering, Rzesz\'ow University of Technology, al.~Powsta\'nc\'ow Warszawy 6, 35-959 Rzesz\'ow, Poland}
\affiliation{Department of Physics and Astronomy, Michigan State University, East Lansing, Michigan 48824, USA}
\author{Vardan Kaladzhyan} 
\affiliation{Institute de Physique Th\'eorique, CEA/Saclay, Orme des Merisiers, 91190 Gif-sur-Yvette Cedex, France}
\affiliation{Laboratoire de Physique des Solides, CNRS, Univ. Paris-Sud, Universit\'e Paris-Saclay, 91405 Orsay Cedex, France}
\author{Cl\'ement Dutreix}
\affiliation{Univ.~Bordeaux, LOMA, UMR 5798, Talence, France and CNRS, LOMA, UMR 5798, Talence, F-33400, France}
\affiliation{Univ.~Lyon, Ens de Lyon, Univ Claude Bernard, CNRS, Laboratoire de Physique, F-69342 Lyon, France}
\author{Cristina Bena}
\affiliation{Institute de Physique Th\'eorique, CEA/Saclay, Orme des Merisiers, 91190 Gif-sur-Yvette Cedex, France}

\begin{abstract}
The bulk-boundary correspondence establishes a connection between the bulk topological index of an insulator or superconductor, and the number of topologically protected edge bands or states. For topological superconductors in two dimensions the first Chern number is related to the number of protected bands within the bulk energy gap, and is therefore assumed to give the number of Majorana band states in the system. Here we show that this is not necessarily the case. As an example we consider a hexagonal-lattice topological superconductor based on a model of graphene with Rashba spin orbit coupling, proximity induced $s$-wave superconductivity, and a Zeeman magnetic field. We explore the full Chern number phase diagram of this model, extending what is already known about its parity. We then demonstrate that despite the high Chern numbers that can be seen in some phases these do not  strictly always contain Majorana bound states.
\end{abstract}


\date{\today}

\maketitle

\section{Introduction}

The search for Majorana bound states (MBS) in condensed matter systems\cite{Qi2011} has already produced a large volume of theoretical work\cite{Kitaev2001,Fu2008,Fu2009,Sato2009,Lutchyn2010,Oreg2010,Qi2011} and promising, though not conclusive, experiments.\cite{Mourik2012,Deng2012,Das2012,Lee2014,Nadj-Perge2014,Wang2012b,Xu2015,Pawlak2016} A one dimensional (1D) topologically non-trivial superconductor will have Majorana bound states present at its ends. This is the result of the well known bulk-boundary correspondence,\cite{Ryu2002} which relates the topological invariant to the number of topologically protected edge states. Two-dimensional topological superconductors, including for example p-wave or s-wave pairing, have also received a lot of attention,\cite{Fu2008,Potter2010,Mizushima2013,Wang2014,Poyhonen2014,Seroussi2014,Wakatsuki2014,Deng2014,San-Jose2014,Thakurathi2014,Mohanta2014,Sedlmayr2015,Rontynen2016,Kaladzhyan2016a,Yang2016} and including the model we consider here of a system with spin-orbit coupling, in the presence of superconducting proximity and a Zeeman field, as well as other related models.\cite{Kane2005a,Qiao2010,Chamon2012,Black-Schaffer2012,Liu2013a,Dutreix2014,Dutreix2014b,Wang2016,Kaladzhyan2017,Dutreix2017}.
In two dimensions (2D) the bulk-boundary correspondence relates the Chern number, $\nu$, to the number of protected bands connecting the bulk states above and below the gap \cite{Teo2010,Hasan2010,Chiu2016,Chan2017,Chan2017a} arising in a ribbon structure. These bands correspond to protected edge states and it is often assumed that their zero-energy crossing corresponds to the formation of a MBS.\cite{Rontynen2016,Kaladzhyan2016a,Wang2016} Indeed the usual expression of the bulk boundary correspondence is in terms of zero energy modes, rather than protected bands.\cite{Teo2010} However as we will demonstrate this is not necessarily the case, and a topologically protected band does not necessarily contain a MBS. 

As an example we focus on a single-layer-hexagonal topological superconductor. This allows us to easily consider two very different types of boundary, both zig-zag and armchair edges. We calculate explicitly the Chern number and we construct a full phase diagram based on the value of the Chern number, and not only on its parity as has been done previously\cite{Dutreix2017} We identify the band structures corresponding to each value of the Chern number and we confirm that it is equal to the number of protected edge bands. We show that the Chern number can be changed by gap closings at many points in the Brillouin zone (BZ), however only gap closings at specific points in the BZ can lead to the formation of protected MBS. The Chern number itself varies from -5 to 5. The number of MBS on an edge can vary however from 0 to 3, rather than 0 to 5, and depends on the type of nanoribbon one is considering. One recent work reported MBS near the Dirac points of this model,\cite{Wang2016} in a phase with $\nu=4$. Such states were not found in other studies.\cite{Chamon2012,Black-Schaffer2012,Dutreix2014,Dutreix2014b,Sedlmayr2015,Dutreix2017} Here we will clarify that although there are protected bands in this phase, strictly speaking there are no MBS present in the lattice model in this phase. 

To understand why this is the case we will introduce a more careful definition of MBS. Thus we show that one must additionally consider if the states found near zero energy scale exponentially to zero in the thermodynamic limit. Also we test the low-energy states in terms of their Majorana-like properties; to this end we use the Majorana polarization,\cite{Sedlmayr2015b,Sticlet2012,Sedlmayr2016} a direct local check of the Majorana nature of an eigenstate. Also we provide symmetry arguments, as well as study the specific manner in which the gap closing may occur in order to give rise to a change in the number of MBS. Finally we consider the effects of disorder on the formation of MBS.

This article is organized as follows. In Sec.~\ref{sec_model} we introduce the model of a two-dimensional hexagonal lattice with induced superconducting proximity, spin-orbit coupling and a Zeeman magnetic field perpendicular to the plane. In Sec.~\ref{sec_chern} we  calculate numerically the Chern number for this model. In Sec.~\ref{sec_parity} we  calculate its parity analytically. In Sec.~\ref{sec_mbs} we consider the bulk-boundary correspondence and how the Chern number relates to the formation of topologically protected bands and of MBS. 
We conclude in Sec.~\ref{sec_conclusions}.

\section{Model}\label{sec_model}

The model under consideration is a hexagonal-lattice topological superconductor that can be realized in graphene with Rashba spin orbit coupling $\alpha$, proximity induced $s$-wave superconductivity $\Delta$, and a Zeeman magnetic potential $B$. We define $t$ as the strength of the nearest-neighbor hopping, and $\mu$ is the chemical potential. Then the Hamiltonian can be written as
\begin{eqnarray}\label{graphene}
H&=&-t\sum_{\substack{<i,j>\\\sigma}}c^\dagger_{i\sigma}c_{j\sigma}-\im\alpha\sum_{\substack{<i,j>\\\sigma,\sigma'}}(\vec{\delta}_{ij}\times{\vec \sigma}_{\sigma\sigma'})_z c^\dagger_{i\sigma}c_{j\sigma'}\\\nonumber
&&+\sum_{i,\sigma}c^\dagger_{i\sigma}\left[B\sigma^z_{\sigma\sigma}-\mu\right]c_{i\sigma}
+\Delta\sum_{i}\left[c^\dagger_{i\uparrow}c^\dagger_{i\downarrow}+c_{i\downarrow}c_{i\uparrow}\right]\,.
\end{eqnarray}
where $c^{(\dagger)}_{i\sigma}$ denotes the annihilation (creation) operator of an electron  of spin $\sigma$ at site $i$, and $\vec\delta_{ij}$ are the nearest neighbor vectors.
This model has been carefully studied both in the presence of a superconducting proximity effect\cite{Dutreix2014,Dutreix2014b,Wang2016,Kaladzhyan2017,Dutreix2017}, as well as without it\cite{Kane2005a,Qiao2010}. We will define the lattice spacing $a=1$ and $\hbar=1$ throughout. The lattice has a length of $L_\parallel$ unit cells along the nanoribbon direction, which always has periodic boundary conditions; and a width of $L_w$ unit cells, which can have periodic or open boundary conditions in the numerical simulations. We are interested in nanoribbons with open edges aligned along both the armchair and zigzag directions of the hexagonal lattice.

\section{Numerical calculation of the Chern number}\label{sec_chern}

In Eq.~(7) in Ref.~\onlinecite{Wang2016} the authors give their results for the Chern number for this model, and show that it can reach values of up to $4$, in one region of phase space. They note that, the gap closings at the Dirac points $\vec k=(\pm4\pi/3\sqrt{3},0)$ in the BZ need to be taken into account, in addition to those at the time reversal invariant (TRI) $\Gamma$ points, $\Gamma_0=(0,0)$, $\Gamma_1=(0,2\pi/3)$, $\Gamma_2=(\pi/\sqrt{3},\pi/3)$, and $\Gamma_3=(\pi/\sqrt{3},-\pi/3)$, which were already known .\cite{Dutreix2014,Dutreix2014b} Here we will show explicitly what the topological phase diagram looks like. In fact there are other points in the BZ where a gap closing changes the Chern number.

\begin{figure}
\includegraphics[width=0.48\columnwidth]{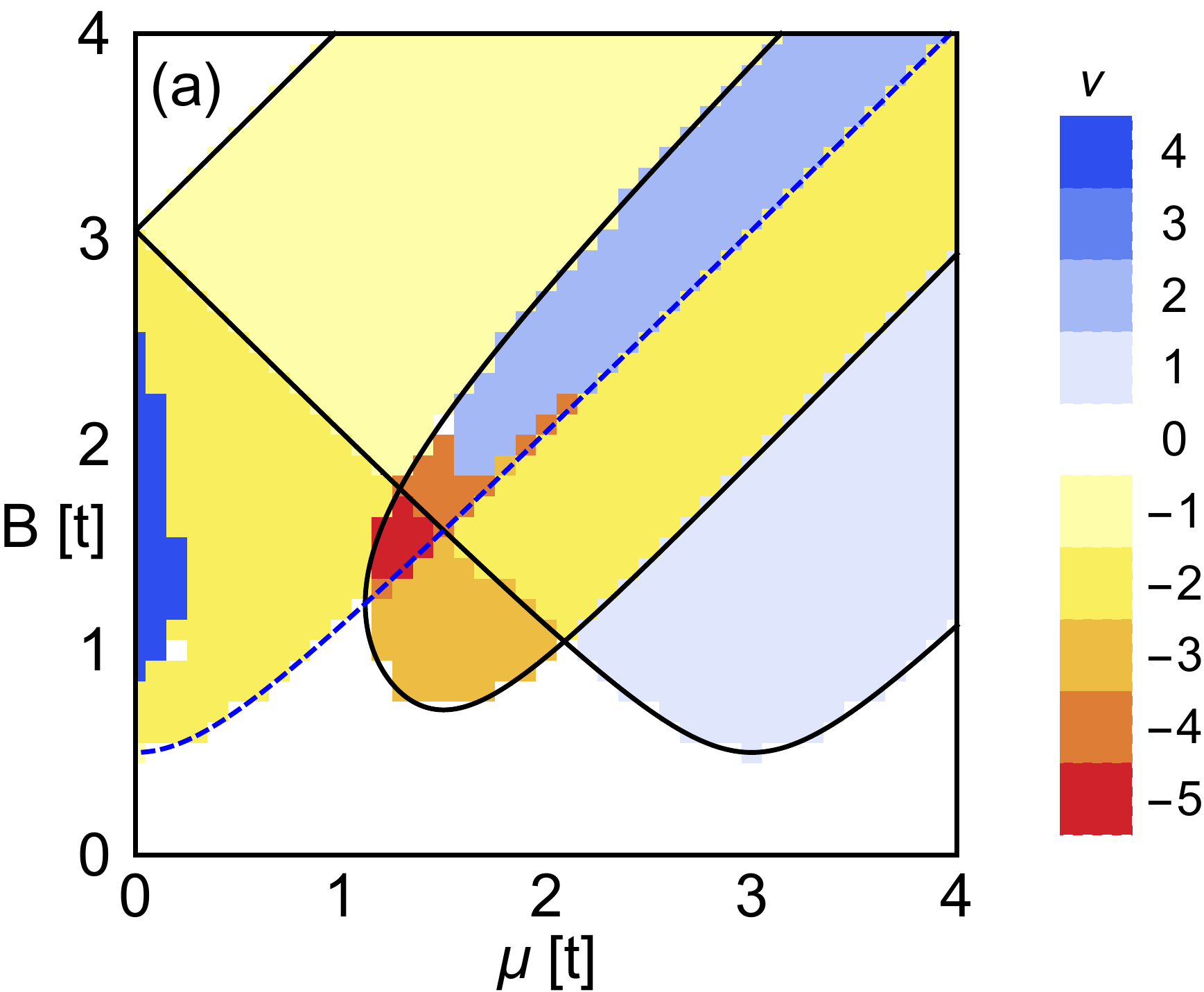}
\includegraphics[width=0.48\columnwidth]{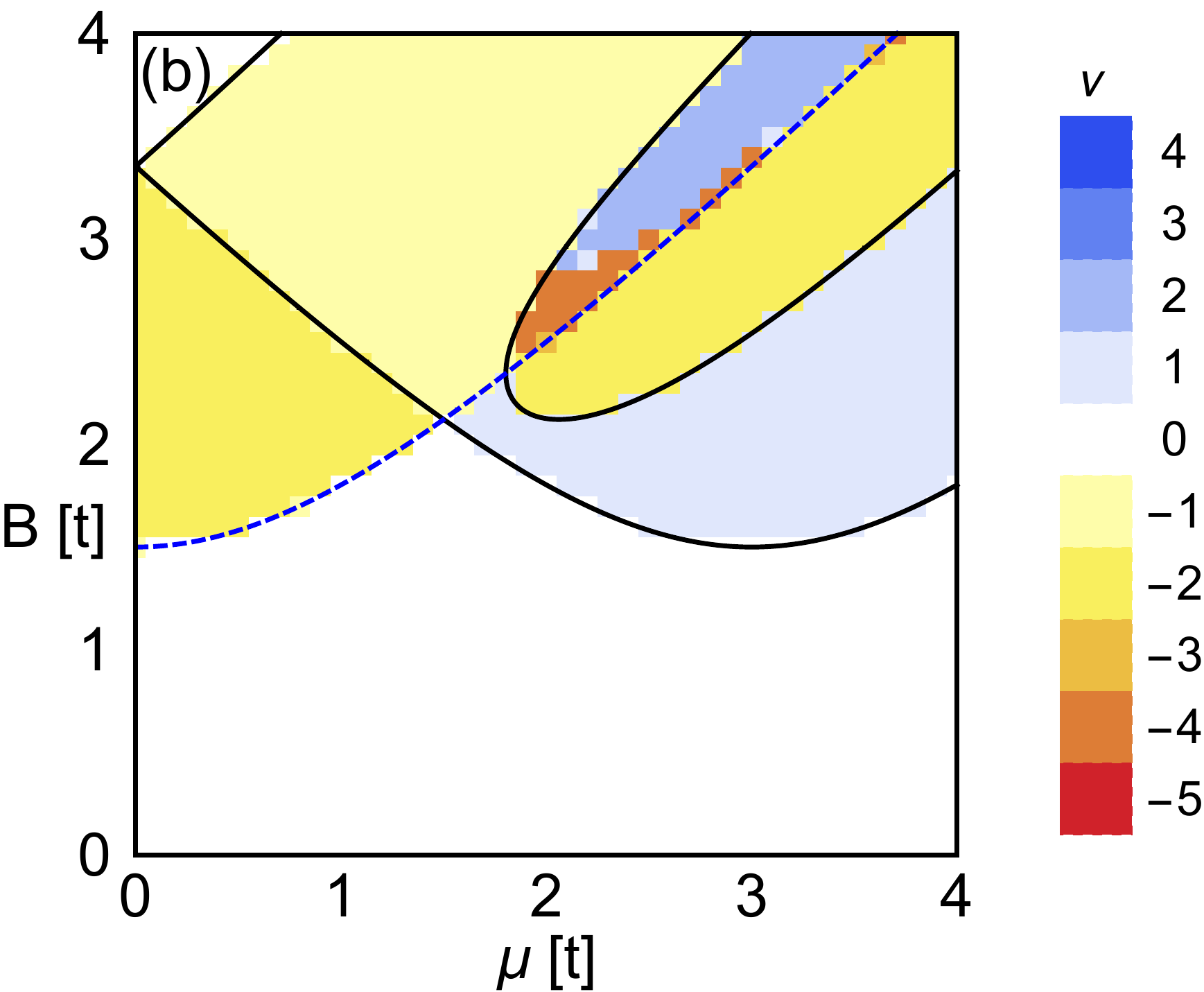}\\
\includegraphics[width=0.48\columnwidth]{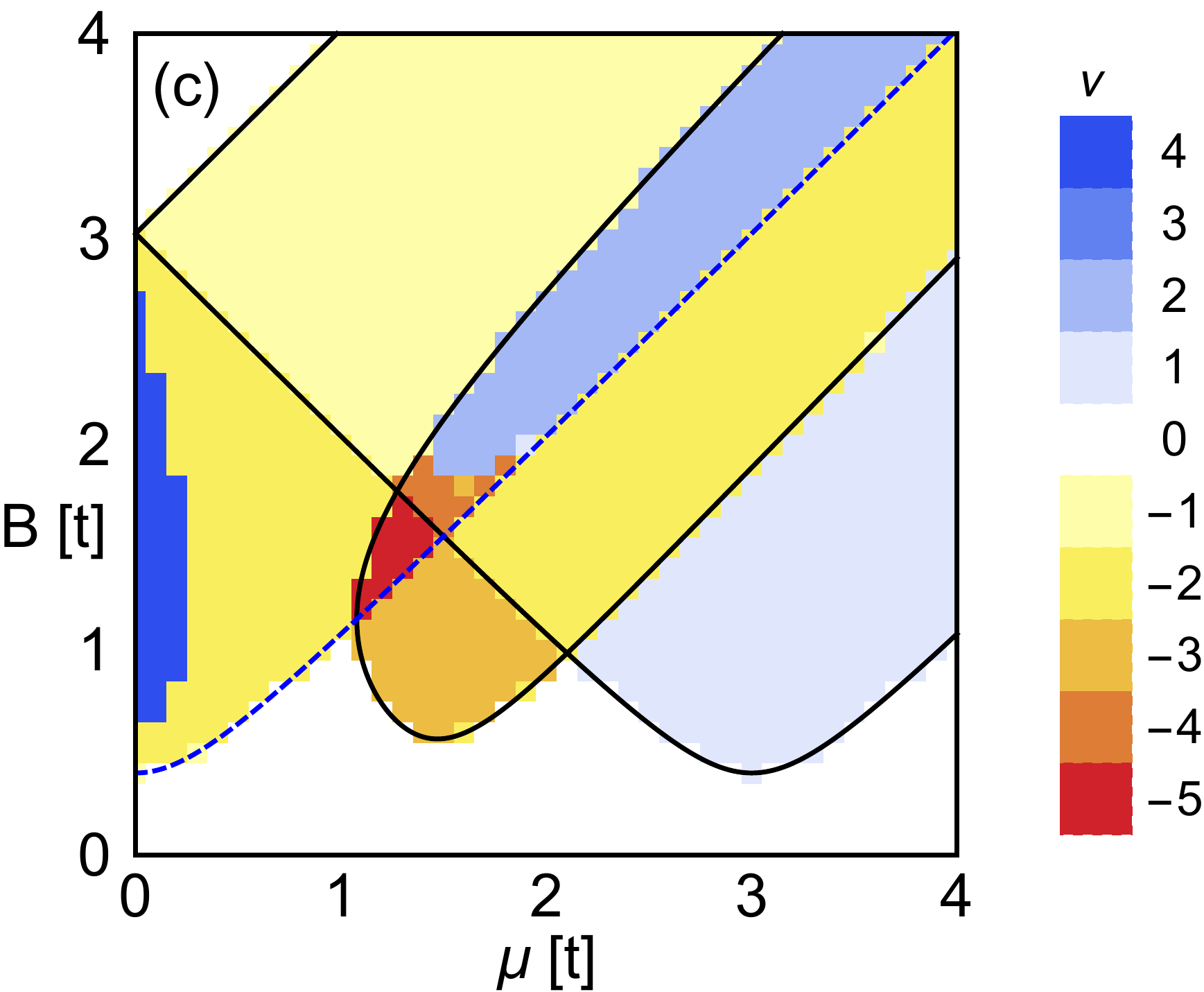}
\includegraphics[width=0.48\columnwidth]{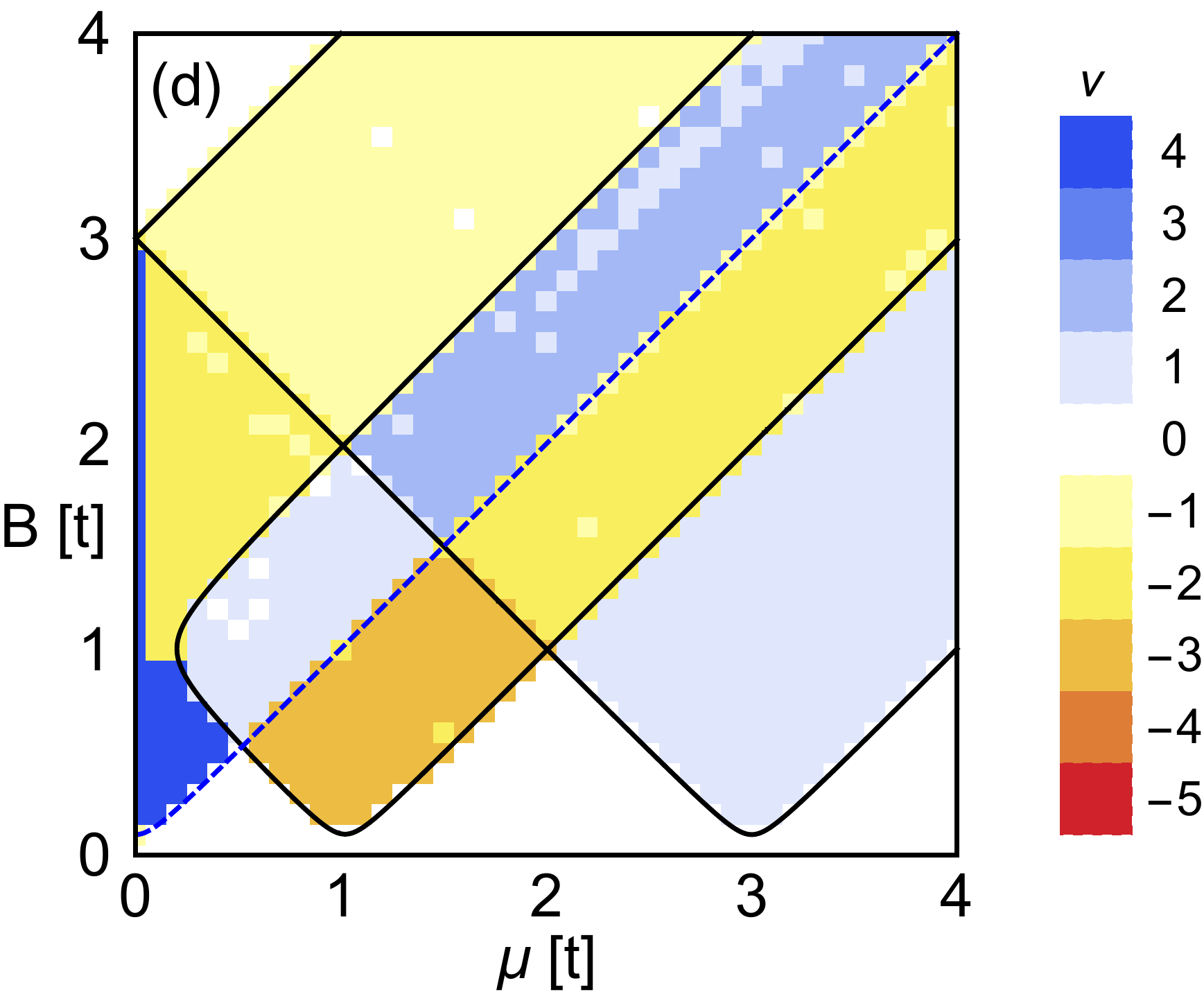}
\caption{(Color online) Numerically determined topological phase diagrams for Eq.~\eqref{graphene} using Eq.~\eqref{tknn}, where $\nu$ is the Chern number. The parameters are: (a) $\alpha=\Delta=0.5t$; (b) $3\alpha=\Delta=1.5t$; (c) $\alpha=0.5t$ and $\Delta=0.4t$; and (d) $\alpha=\Delta=0.1t$. Solid black lines show the phase boundaries caused by the gap closings at the TRI momenta and the blue dashed line corresponds to the gap closes at the Dirac point. These are not however the only phase boundaries.}
\label{numchern}
\end{figure}

Following Ref.~\onlinecite{Ghosh2010} the Chern number, or equivalently the Thouless-Kohmoto-Nightingale-den Nijs (TKNN) invariant\cite{Thouless1982}, can be calculated numerically using
\begin{eqnarray}\label{tknn}
\nu=\frac{\im}{8\pi^2}\int\ud^2 k\ud\omega\Tr&&\big[G^2(\partial_{k_y}H)G(\partial_{k_x}H)\\\nonumber&&\qquad-G^2(\partial_{k_x}H)G(\partial_{k_y}H)\big]\,,
\end{eqnarray}
where $G=(H-\im\omega)^{-1}$ and $\vec{k}=(k_x,k_y)$ is the momentum. This can be implemented numerically, and some examples are given in Fig.~\ref{numchern}. The Chern number can be as large as -5, which is unusually high for such a model.

\begin{figure}
\includegraphics[width=0.68\columnwidth]{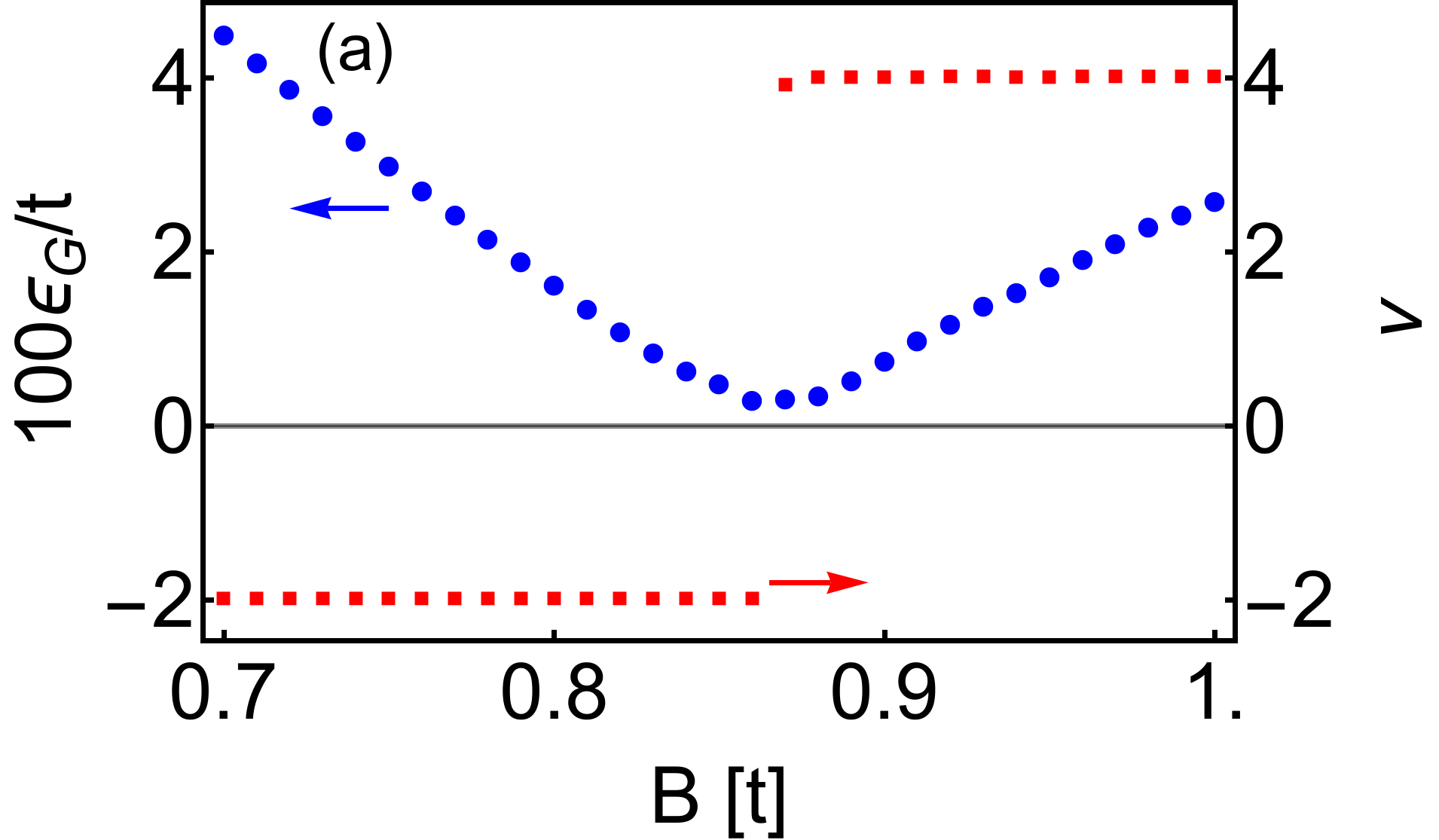}\\
\includegraphics[width=0.68\columnwidth]{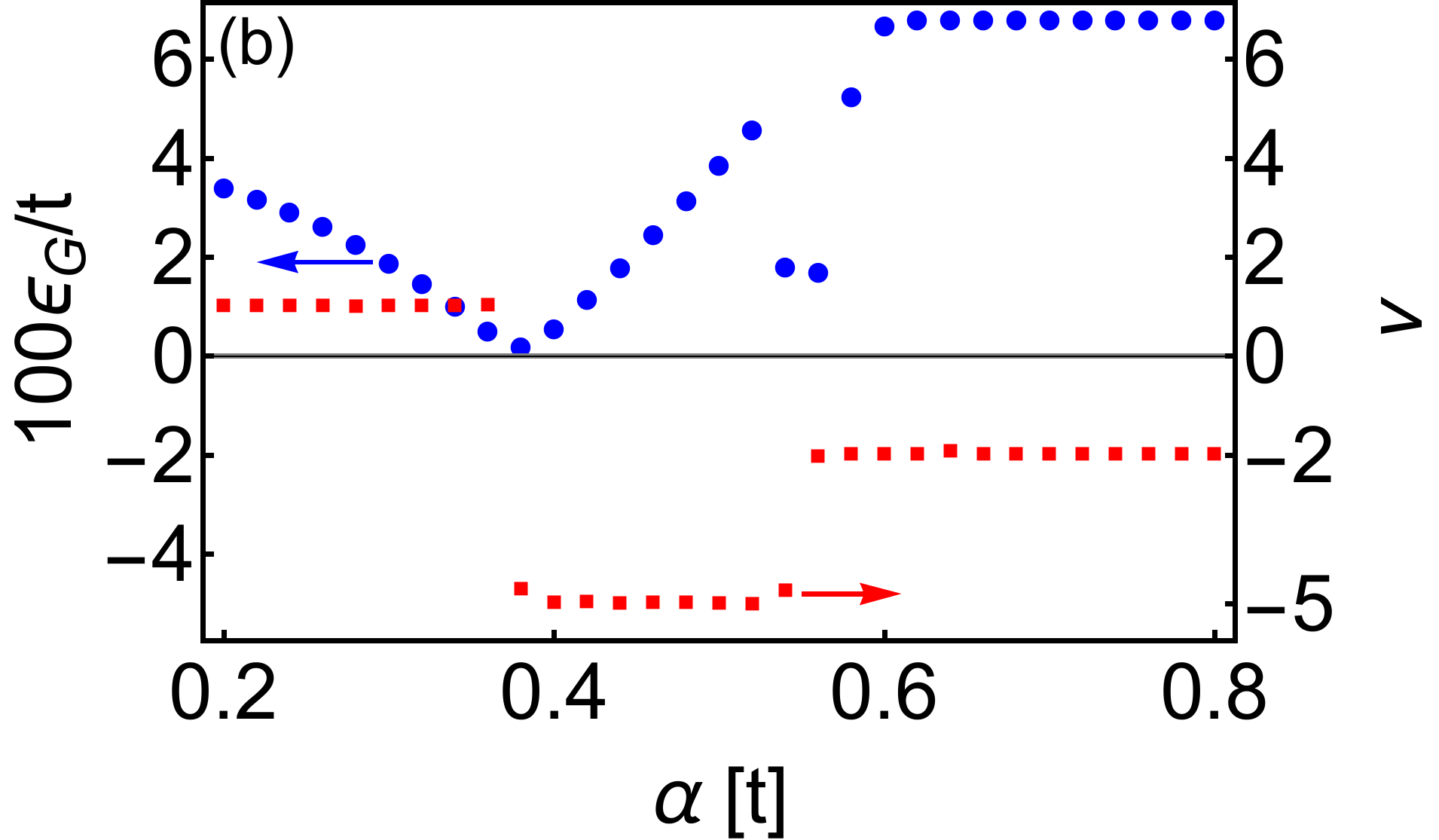}
\caption{(Color online) Numerically determined topological phase diagrams for Eq.~\eqref{graphene} using Eq.~\eqref{tknn}, where $\nu$ is the Chern number (red squares), compared to the gap closing (blue circles) where $\epsilon_G$ is the gap. (a) shows a transition from $\nu=-2\to4$ as a function of $B$ (accompanied by a gap closing at the Dirac point) with the gap calculated for a system of length and width $L=L_w=L_\parallel=500$. The parameters are $\Delta=\alpha=0.5t$ and $\mu=0$. This is a cut through the phase diagram in Fig.~\ref{numchern}(a). (b) shows transitions through $\nu=1\to-5\to-2$ as a function of $\alpha$ (accompanied by gap closings at points in the BZ) with the gap calculated for $L=500$ as for (a). The parameters are $\Delta=0.6t$ $B=1.5t$ and $\mu=1.3t$. }
\label{numcherngap}
\end{figure}

The phase diagrams in Fig.~\ref{numchern} show changes in Chern numbers away from the analytically calculated bulk gap closing lines for the $\Gamma$ and Dirac points. To be certain that this is not a numerical error we track the bulk gap across some of these transitions, see Fig.~\ref{numcherngap}. For all the changes in the Chern number we can see that the gap does close at some point in the BZ as required. One can also explicitly check that the bulk-boundary correspondence holds, demonstrating that these regions are not caused by numerical errors. However, as we shall see in what follows, these high Chern numbers do not necessarily lead to large numbers of protected MBS.

In Fig.~\ref{numchern} it can be observed that the gap closings at the $\Gamma$ points alter the Chern number from odd to even or vice versa. In the following section we will demonstrate why this occurs.

We note that this model can still be a topological insulator when $\Delta=0$, though this is by no means guaranteed. However there is no general connection between the protected edge modes we see here and those in the topological insulator.  In some cases the edge modes persist and remain topologically protected, but this is certainly not a generic feature. In general as $\Delta\to0$ the gap can close and reopen, changing the topology, and some gapped phases here simply become ungapped metallic phases in this limit.

\section{Parity of the Chern number}\label{sec_parity}

A detailed calculation of the parity of the Chern number for Eq.~\eqref{graphene} has been performed and presented in detail in Refs.~\onlinecite{Dutreix2017} and \onlinecite{Dutreix2014b}, and an example is available also in the appendix of Ref.~\onlinecite{Sedlmayr2015}.
We show that a relation between the Chern number and the parity of the bands at the TRI momenta, $\Gamma_0=(0,0)$, $\Gamma_1=(0,2\pi/3)$, $\Gamma_2=(\pi/\sqrt{3},\pi/3)$, and $\Gamma_3=(\pi/\sqrt{3},-\pi/3)$, can be proven for Eq.~\eqref{graphene}. Here we have set the lattice spacing $a=1$. The $K=(4\pi/3\sqrt{3},0)$ and $K'=(-4\pi/3\sqrt{3},0)$ Dirac points do not affect the parity of the Chern number, and neither do any of the other gap closing points. This is due to the fact that they always paired,  with the gaps at $\vec k$ and $-\vec k$ necessarily closing at the same time, thus the {\it parity} of the Chern number is only altered by the special $\Gamma$ points. However, as we have seen, the Chern number itself can be changed by gap closings at various points in the BZ. 

In momentum space the total Hamiltonian \eqref{graphene} is
\begin{equation}
H=\frac{1}{2}\sum_{\vec k}\Psi^\dagger_{\vec k}\mathcal{H}(\vec k)\Psi_{\vec k}
\end{equation}
with the fermion field $\Psi^\dagger_{\vec k}=(a^\dagger_{{\vec k}\uparrow},b^\dagger_{{\vec k}\uparrow},a^\dagger_{{\vec k}\downarrow},b^\dagger_{{\vec k}\downarrow},a_{-{\vec k}\uparrow},b_{-{\vec k}\uparrow},a_{-{\vec k}\downarrow},b_{-{\vec k}\downarrow})$. Here $a^\dagger_{{\vec k}\sigma}$ and $b^\dagger_{{\vec k}\sigma}$ create electrons of spin $\sigma$ and with momentum $\vec{k}$ on the two sublattices. The Hamiltonian matrix is
\begin{equation}
\mathcal{H}(\vec k)=\begin{pmatrix}
H(\vec k)&\Delta(\vec k)\\
-\Delta^*(\vec k)&-H^*(-\vec k)
\end{pmatrix}\,,
\end{equation}
where each entry is itself a $4 \times 4$ matrix. The pairing matrix satisfies $\Delta^*(-\vec{k})=\Delta(\vec{k})$ and the Hamiltonian matrix satisfies particle hole symmetry:
\begin{equation}
\mathcal{C}^\dagger\mathcal{H}(\vec k)\mathcal{C}=-\mathcal{H}^*(-\vec k)\,,
\end{equation}
where $\mathcal{C}=\sigma^0\otimes\lambda^0\otimes\tau^x$. We will use $\sigma$, $\lambda$, and $\tau$ Pauli matrices for the spin, sublattice, and particle-hole sectors and $\sigma^0=\lambda^0=\tau^0\mathbb{I}_2$, the $2 \times 2$ identity matrix. One consequence of the particle hole symmetry is that all the non-zero energy states are paired with a state of opposite energy.

The topological invariant can be related to a parity-like operator, $P=\sigma^x\otimes\lambda^z\otimes\tau^z$, of the negative energy bands at the time reversal invariant momenta.\cite{Sato2009a} All eigenstates at the points, $\Psi_n(\Gamma_i)$, have a definite parity $\pi_n(\Gamma_i)=\pm1$.\cite{Dutreix2017} Note that a sign change in
\begin{equation}
\delta_i=\prod_{E_n<0}\pi_n(\Gamma_i)\,,
\end{equation}
where $E_n$ is the eigenenergy of the eigenstates $\Psi_n(\Gamma_i)$, implies a gap closing at zero energy. 

The Chern number can be defined as the integral of the Berry curvature over the Brillouin zone for the negative energy bands
\begin{equation}
\nu=\frac{1}{2\pi}\int\ud^2k\,\nabla_{\vec k}\times\mathbf{A}^-(\vec k)
\end{equation}
Where the Berry connection is
\begin{equation}
\mathbf{A}^-(\vec k)=\im\sum_n\langle\Psi_n(\vec k)|\nabla_{\vec k}|\Psi_n(\vec k)\rangle=\frac{\im}{2}\nabla_{\vec k}\ln\Det M(\vec k)\,,
\end{equation}
with $M_{mn}=\langle\Psi_m(\vec k)|P\mathcal{C}|\Psi_n(\vec k)\rangle$ and $|\Psi_n(\vec k)\rangle$ an eigenstate of the Hamiltonian. Thus one finds
\begin{equation}
(-1)^\nu=\prod_{E_n<0}\pi_n(\Gamma_0)\pi_n(\Gamma_1)\pi_n(\Gamma_2)\pi_n(\Gamma_3)
\end{equation}
for the topological invariant.

\begin{figure}
\includegraphics[width=0.49\linewidth]{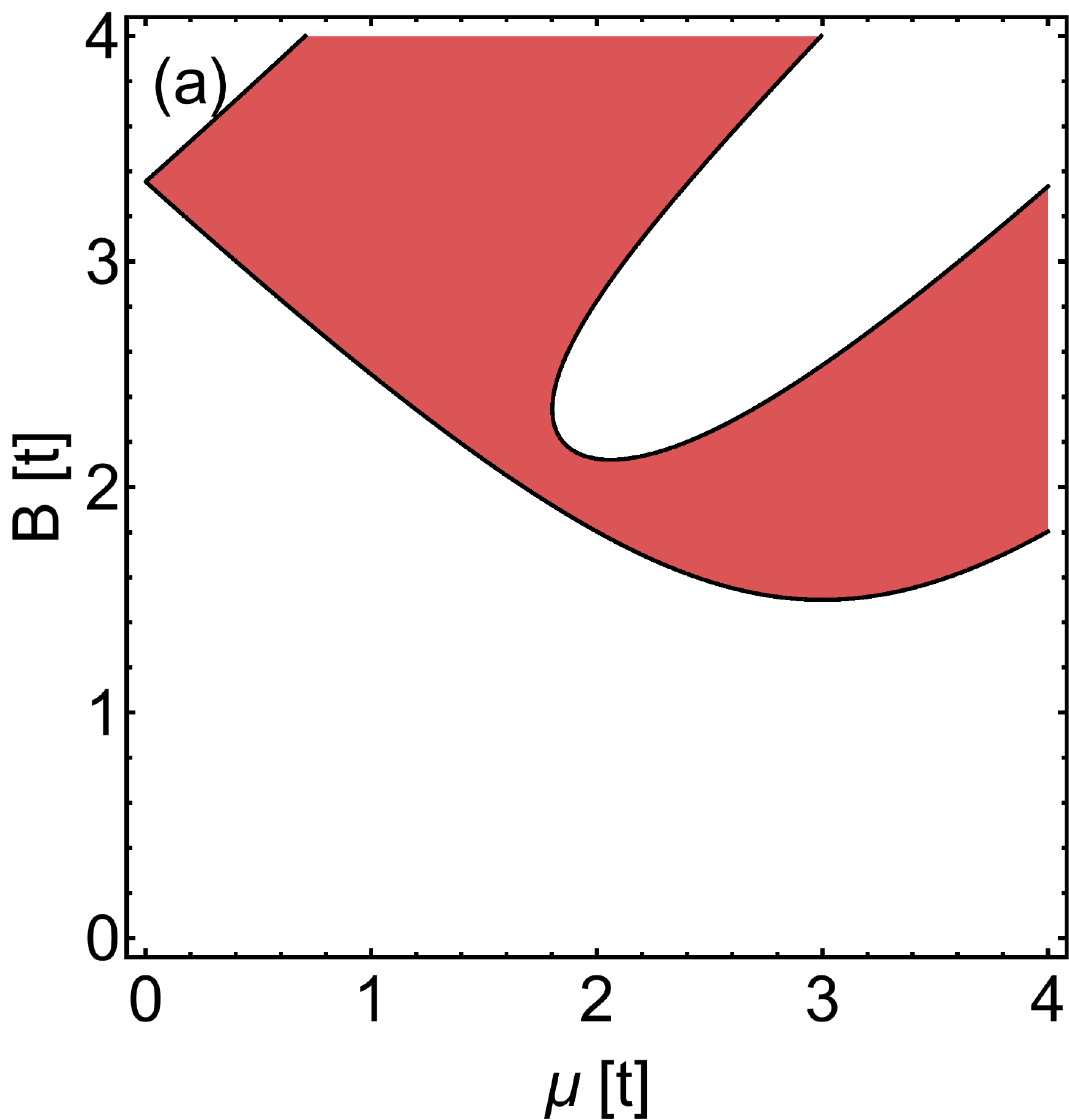}
\includegraphics[width=0.49\linewidth]{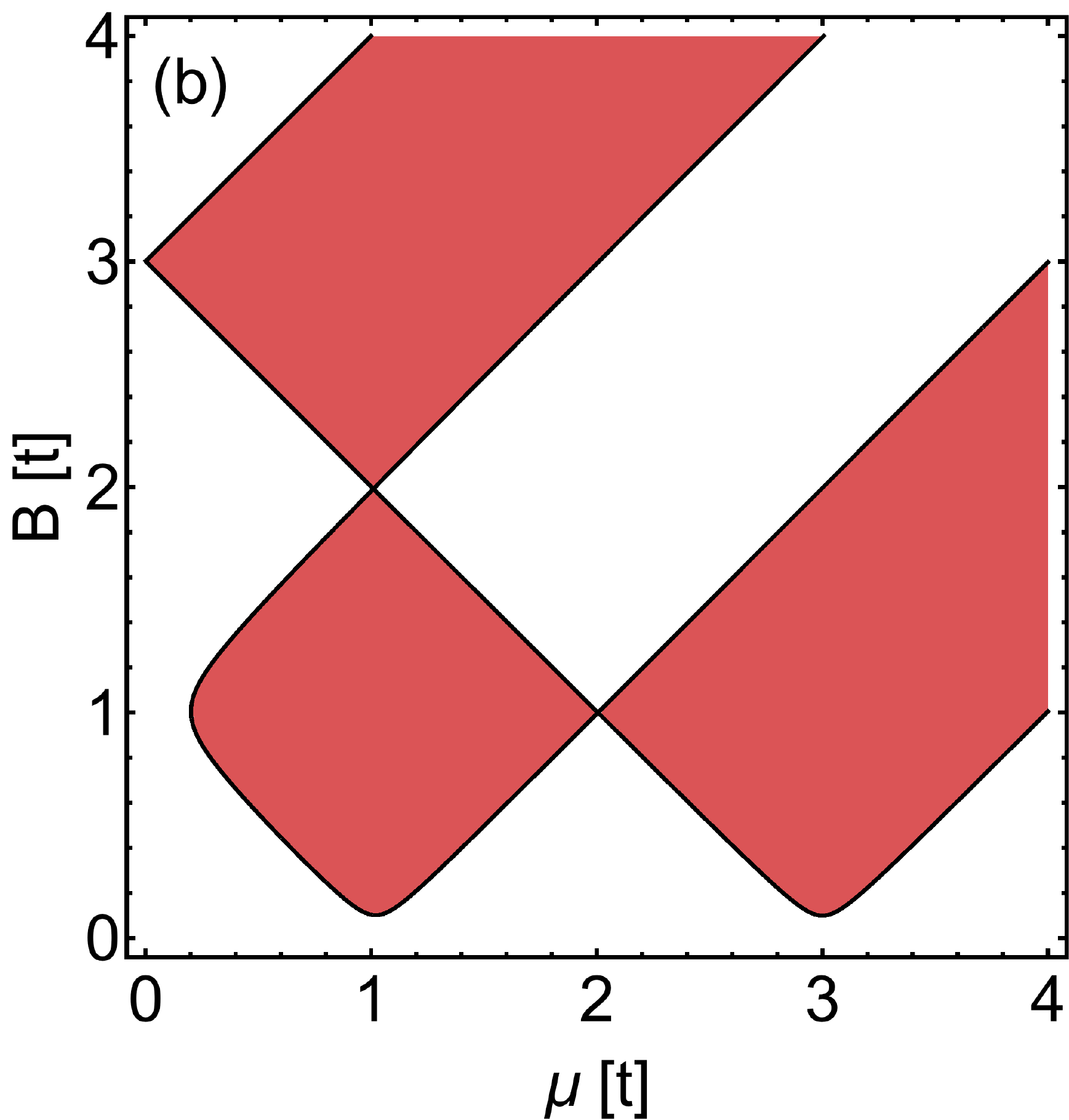}
\caption{(Color online) Topological phase diagrams for Eq.~\eqref{graphene}. The parameters are (a) $3\alpha=\Delta=1.5t$, and (b) $\alpha=\Delta=0.1t$. Compare with Fig.~\ref{numchern}(b,d) respectively. The red regions are satisfy $(-1)^\nu=-1$ and the white regions satisfy $(-1)^\nu=1$, the solid black lines show the phase boundaries and $\nu$ is the Chern number.}
\label{PD}
\end{figure}

This results in\cite{Dutreix2017}:
\begin{widetext}
\begin{eqnarray}
(-1)^\nu&=&\sgn\big[\left(B^4+9^2t^4+18t^2[\Delta^2-\mu^2]+[\Delta^2+\mu^2]^2-2B^2(9t^2+\Delta^2+\mu^2)\right)\\\nonumber
&&\times\left(B^4+t^4+16\alpha^4+8\alpha^2[\Delta^2-\mu^2]+\Delta^4+2\Delta^2\mu^2+\mu^4+2t^2[4\alpha^2+\Delta^2-\mu^2]-2B^2[t^2-4\alpha^2+\Delta^2+\mu^2]\right)\big]\,.
\end{eqnarray}
\end{widetext}
Two exemplary phase diagrams as a function of magnetic field and chemical potential are shown in Fig.~\ref{PD}, which are consistent with the results in Fig.~\ref{numchern}.

\section{Nanoribbon bandstructures and Majorana bound states}\label{sec_mbs}

\subsection{Correspondence between the Chern number and the band structure}
Before we consider the formation of MBS, we will demonstrate in what way the bulk-boundary correspondence manifests itself in this system. The nanoribbons we consider are periodic in one direction and finite with open boundary conditions in the perpendicular direction. We define $k_\parallel \in[-\pi,\pi)$ as the momentum parallel to the edges. We will consider several examples of ribbons, with both zigzag and armchair edges.

\begin{figure}
\includegraphics[width=0.48\linewidth]{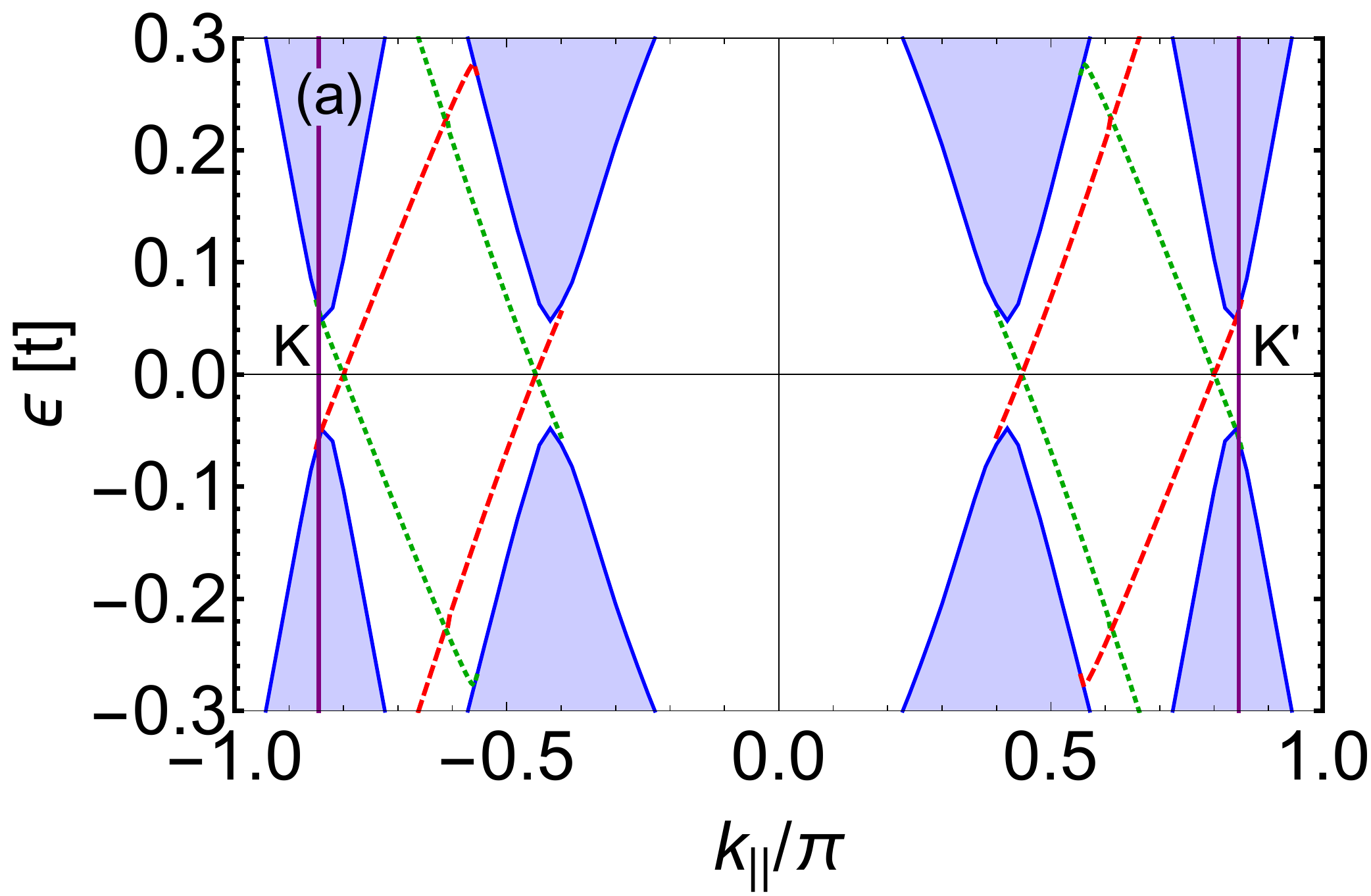}
\includegraphics[width=0.48\linewidth]{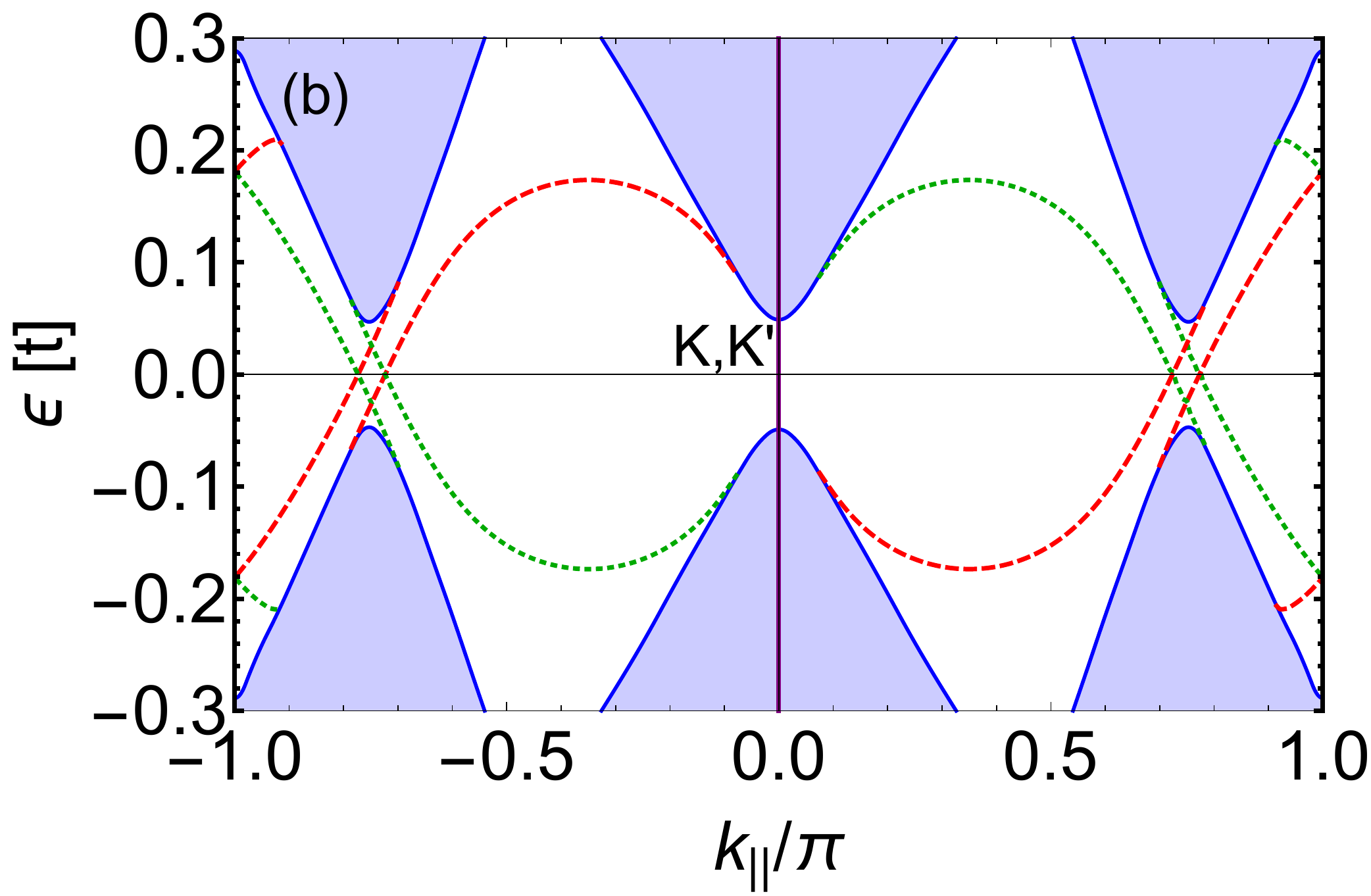}\\
\includegraphics[width=0.48\linewidth]{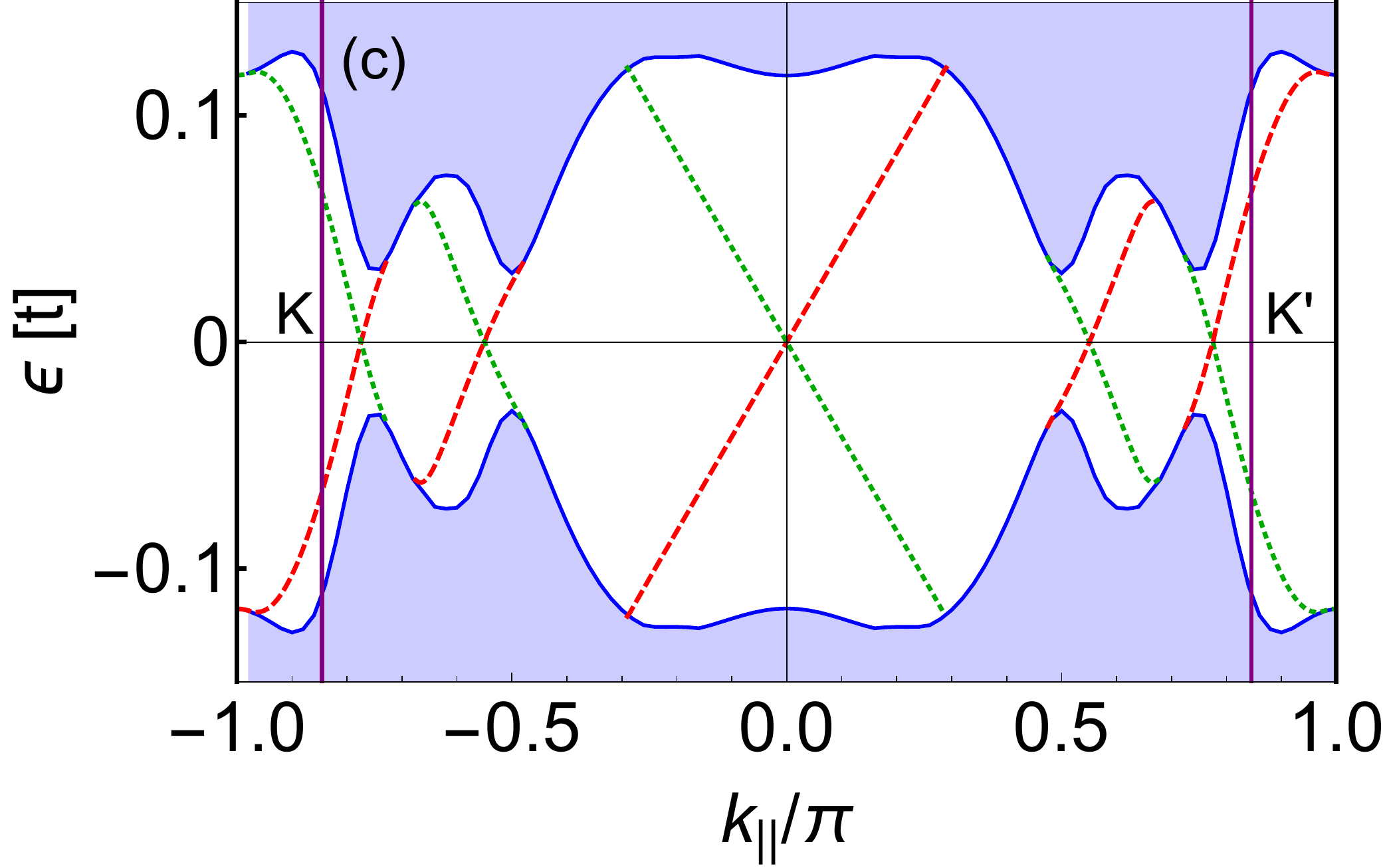}
\includegraphics[width=0.48\linewidth]{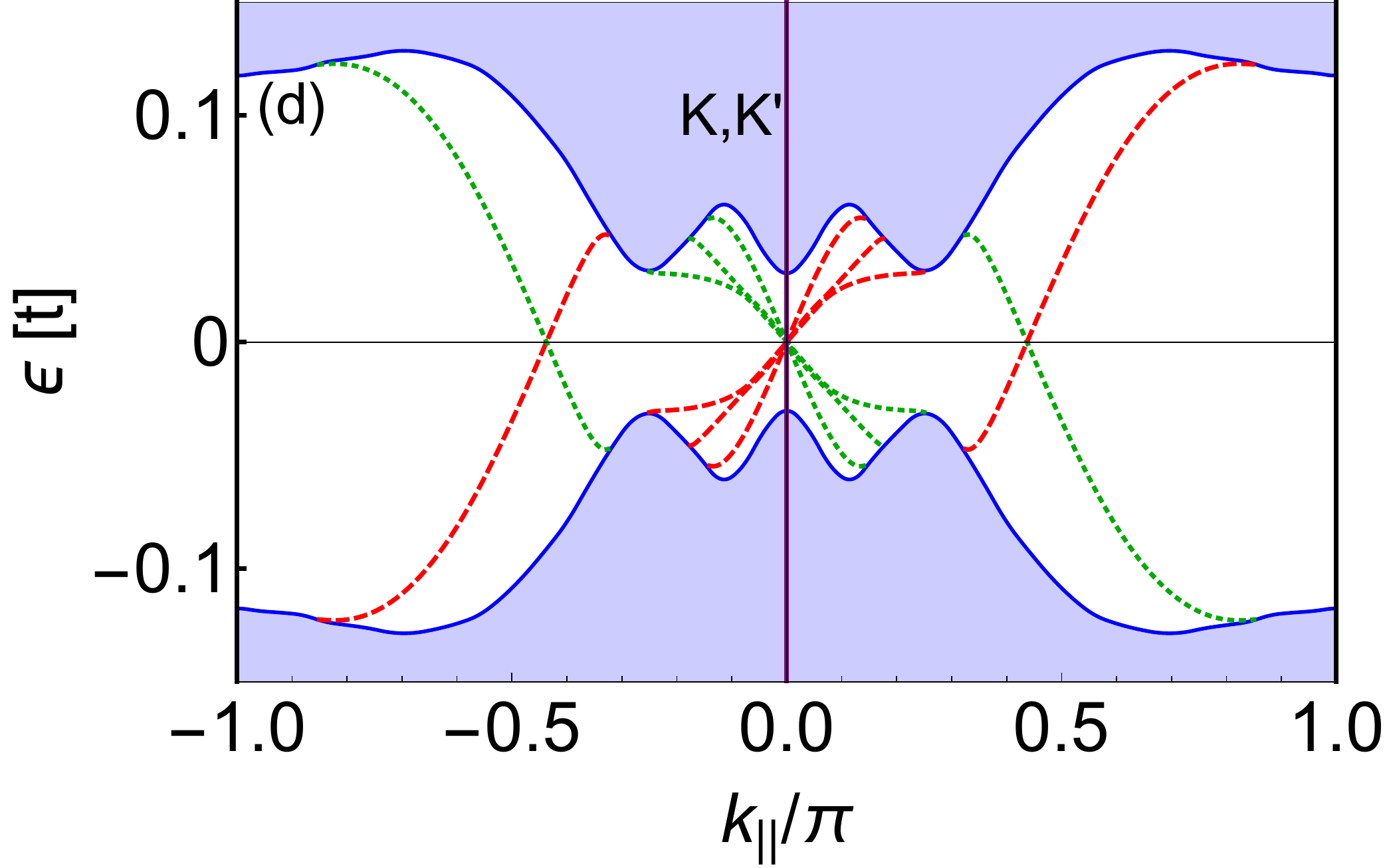}
\caption{(Color online) (a,b) The bandstructures of zigzag (a) and armchair (b) nanoribbons in a regime with $\nu=4$. The parameters are $\alpha=\Delta=0.5t$, $\mu=0.1t$, and $B=1.4t$. (c,d) The bandstructures corresponding to $\nu=-5$ for zigzag (c) and armchair (d) nanoribbons. The parameters are $\alpha=\Delta=0.5t$, $\mu=1.5t$, and $B=1.3t$. In (d) three pairs of edge bands are crossing at $k_\parallel=0$ and there are three MBS per edge in this case. The topologically protected bands localized on one edge are depicted by a dashed red line, while those localized on the other edge are represented by a dotted green line. The K and K' points are marked in the figures for reference.}
\label{BS}
\end{figure}

\begin{figure}
\includegraphics[width=0.48\linewidth]{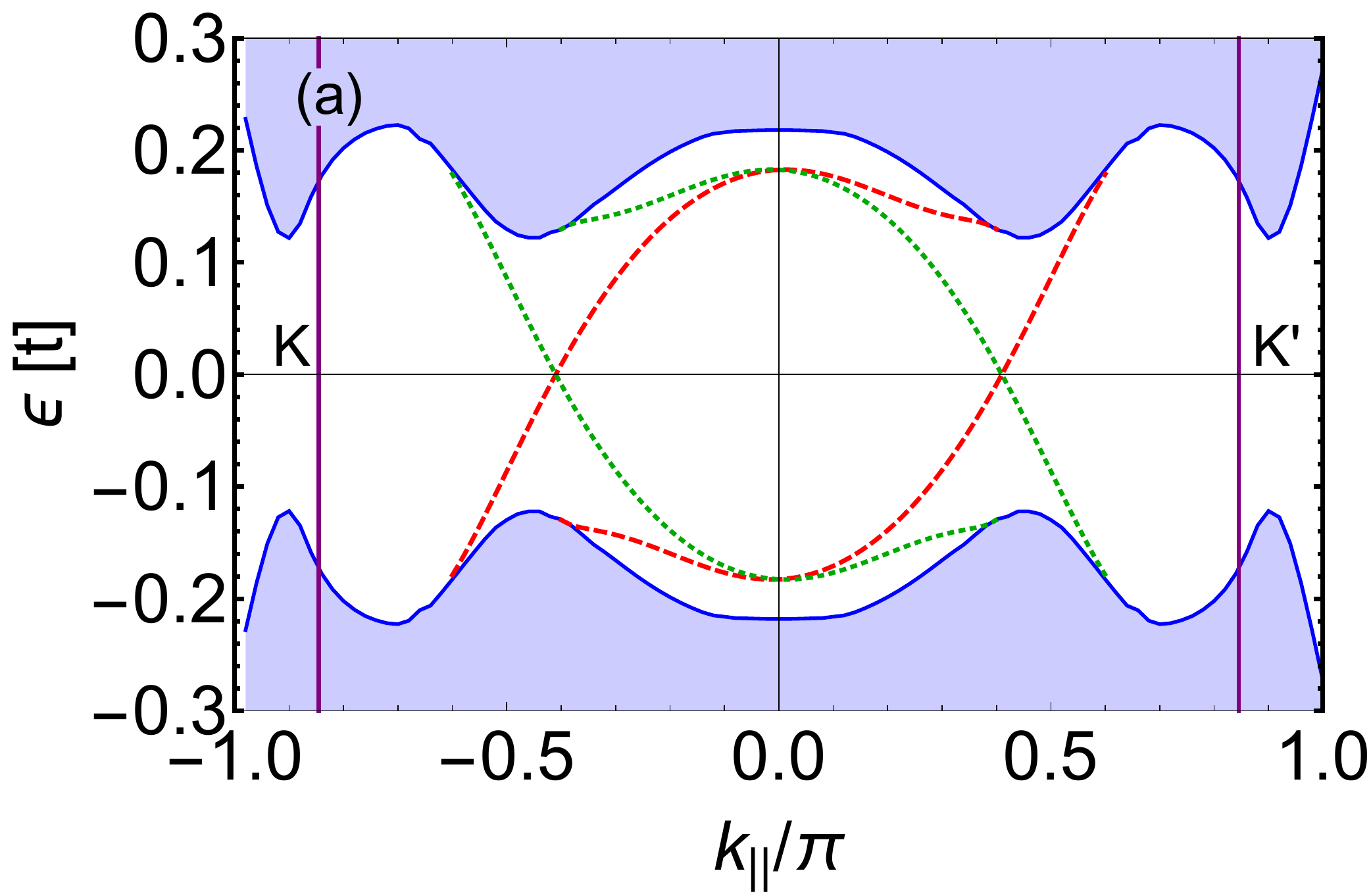}
\includegraphics[width=0.48\linewidth]{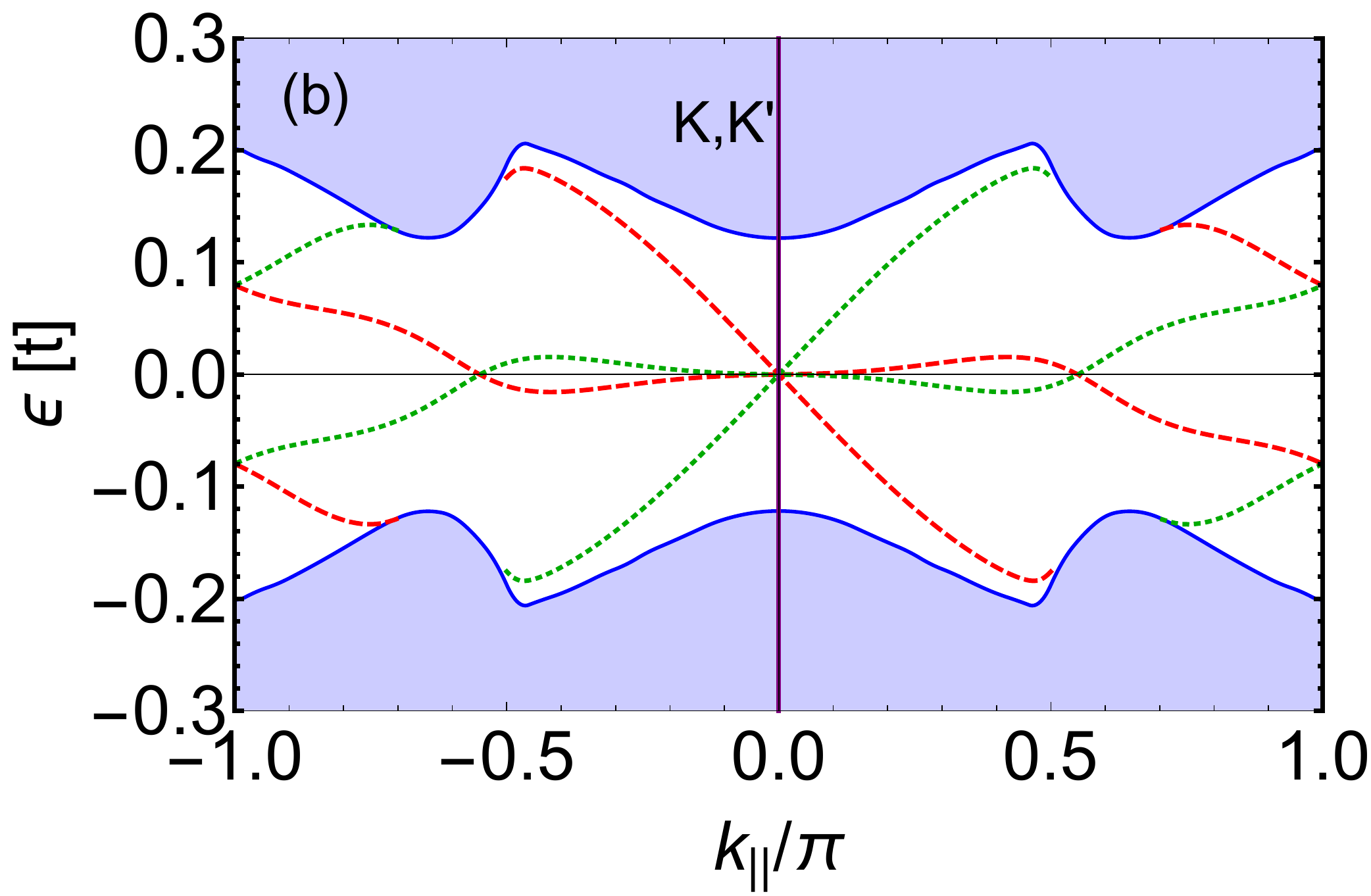}
\caption{(Color online) (a,b) The bandstructure of zigzag (a) and armchair (b) nanoribbons in a regime with $\nu=-2$. The parameters are $\alpha=\Delta=\mu=0.5t$, and $B=1.5t$.  In (b) extra unprotected crossings can be seen at $k_\parallel\neq0$ which can be removed by continuously deforming the bands, see Fig.~\ref{BS_Dis}(a). }
\label{BS2}
\end{figure}

Figs.~\ref{BS}(a,b) correspond to a phase with $\nu=4$. In both cases we observe four pairs of edge bands crossing the bulk gap. The energy of these bands has a monotonic dependence on $k_\parallel$, with the right-moving states being located on one edge, and the left-moving ones on the other. Figs.~\ref{BS}(c,d) correspond to a phase with $\nu=-5$, and we observe five pairs of protected bands. Fig.~\ref{BS2}(a,b) correspond to a phase with $\nu=-2$. Both nanoribbons exhibit two pairs of protected bands crossing zero energy, thus expected to support MBS, however for the armchair nanoribbon, see Fig.~\ref{BS2}(b), one can see that a pair of bands exhibits additional unprotected zero energy crossings close to $k_\parallel=\pm0.5\pi$. Such crossings are unprotected in the sense that a continuous deformation of the band can remove these zero energy crossings, and the corresponding zero energy states are not topologically protected MBS, i.e.~a perturbation can gap out these states. Indeed, this can be seen for example in Fig.~\ref{BS_Dis} where we study the effects of disorder on the band structure.
 
\subsection{Zero-energy states and their identification as MBS based on scaling arguments}
Given the arguments in the previous section, we note that each topologically protected edge band has one protected zero-energy crossing. We would thus naturally expect that each edge band gives rise to a MBS, and thus that the Chern number gives the number of topologically protected MBS. However we will argue in what follows that this is not always the case.

We first note that the band structure is the result of a Fourier transform of Eq.~\eqref{graphene} along the direction parallel to the edge of the ribbon, along which we have imposed periodic boundary conditions. This yields a set of 1D Hamiltonians, $H_{1D}(k_\parallel)$, labelled by the quantum numbers $k_\parallel=2\pi n/L_\parallel$, with $L_\parallel$ being the length of the ribbon in the direction parallel to edges, and $n=0,1,\ldots L_\parallel$. In the thermodynamic limit $k_\parallel$ becomes a continuous variable. In order that a MBS forms, one edge band needs to contain a state with exactly zero energy. While this is of course automatic in the thermodynamic limit, in the finite-size system, when $k_\parallel$ is only taking discrete values, this can only happen at special points in the bandstructure, here for example at $k_\parallel=0,\pi$,  since the energy of a state corresponding to an arbitrary $k_\parallel=0,\pi$ is never exactly zero, but is of the order of $1/L_\parallel$. We propose that the deciding difference between real MBS and non-MBS states lies in how their energy scales to zero in the thermodynamic limit, with the energy of the real MBS decreasing exponentially, while the energy of the non-MBS decreasing inverse proportionally to the system size. 

\begin{figure}
\includegraphics[width=0.48\linewidth]{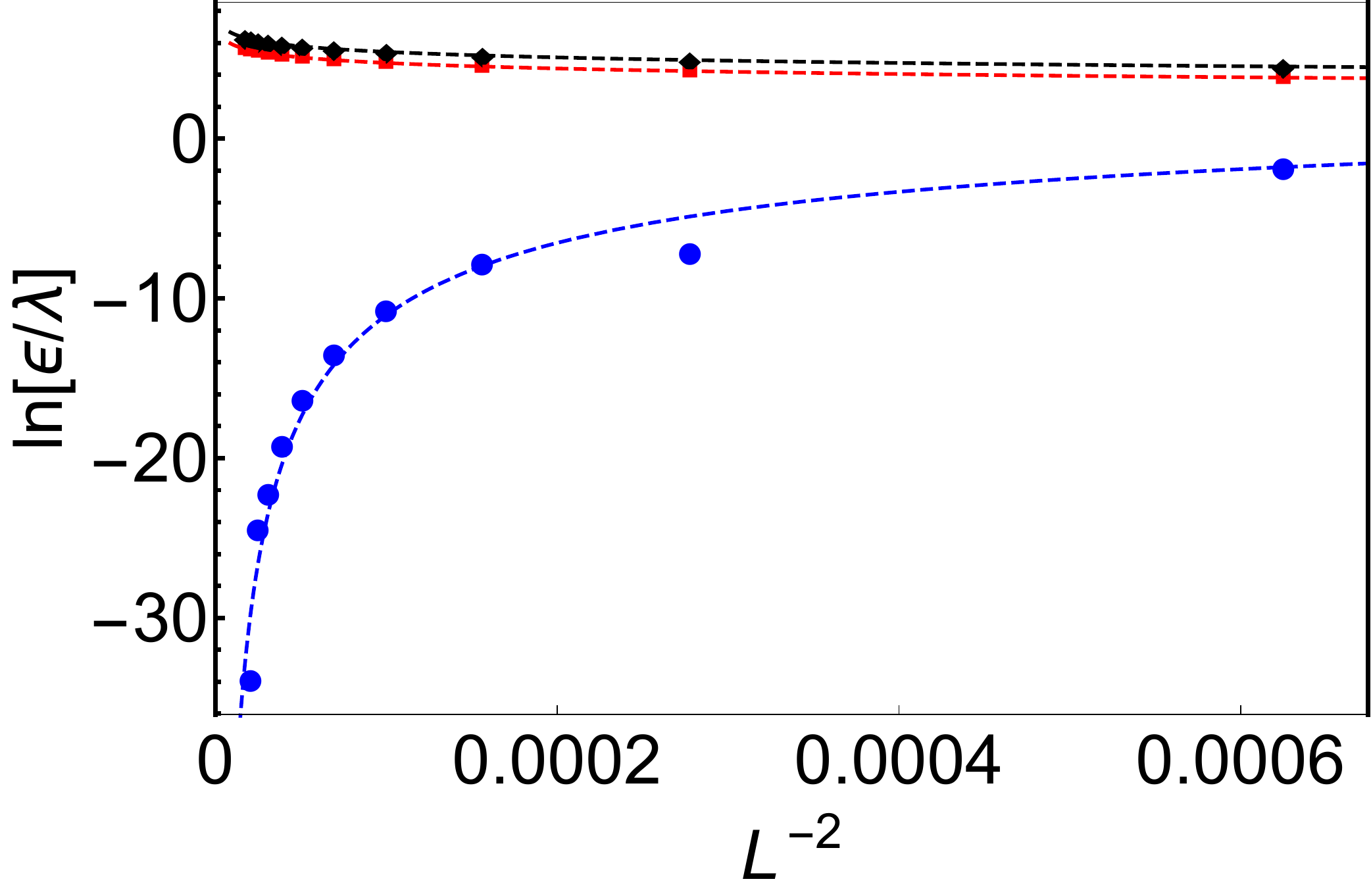}
\includegraphics[width=0.48\linewidth]{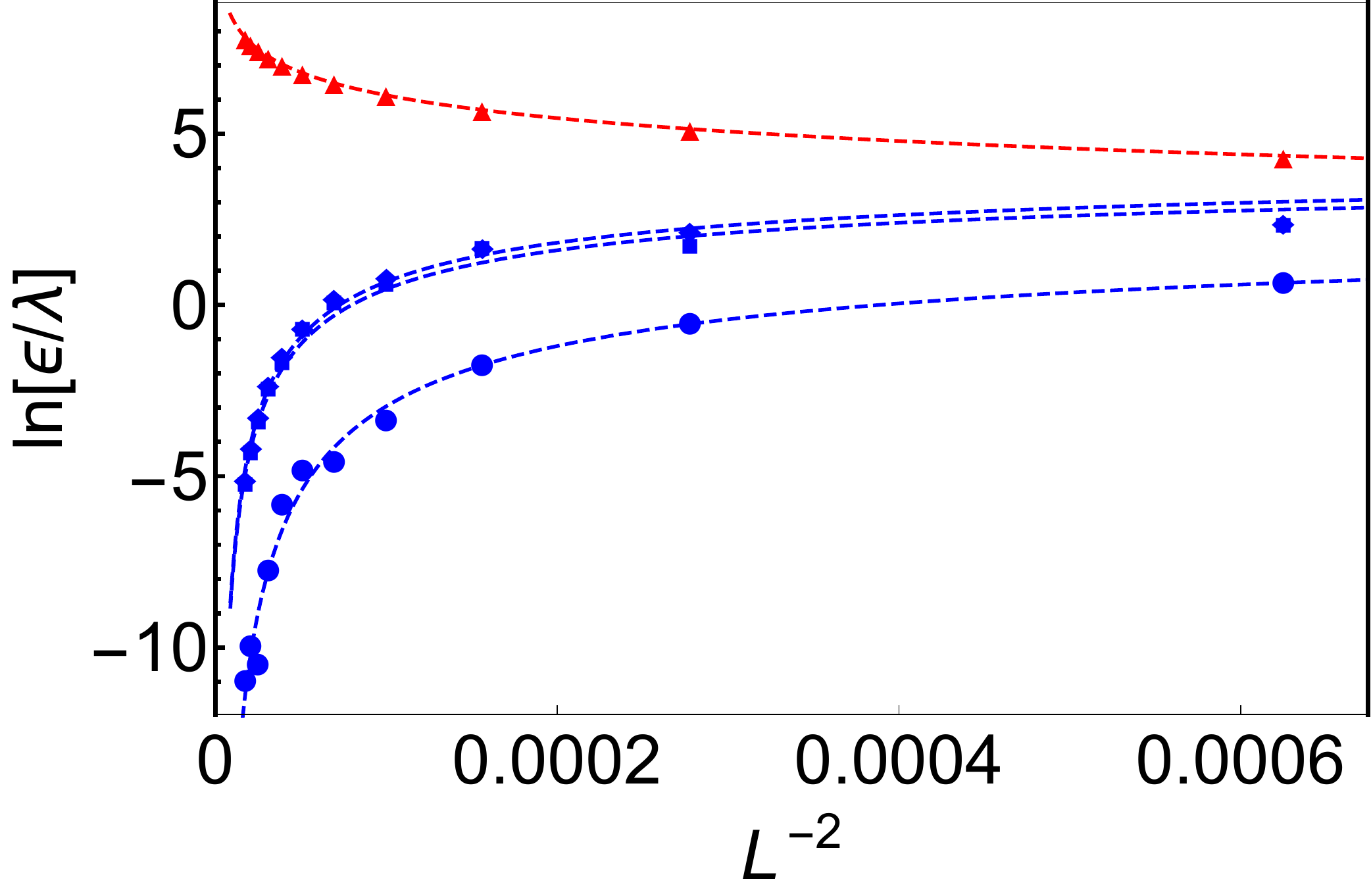}
\caption{(Color online) The energy scaling of the lowest energy state in each band normalized by the mean level spacing, $\epsilon/\lambda$. Here we consider a zigzag nanoribbon (left hand panel) and an armchair nanoribbon (right hand panel) in a $\nu=-5$ phase, same as in Figs.~\ref{BS}(c,d).  We take $L=L_w=L_\parallel$ to be the length and width of system. The low energy states at $k_\parallel=0$ (denoted in blue), show an exponential scaling to zero as $\epsilon/\lambda \sim L^2\e^{-L/\tilde L}$ (dashed blue line), the other low energy states at $k_\parallel\neq0,\pi$, (denoted in red and black) do not scale to zero but show a weaker positive divergence, $\ln \epsilon/\lambda \sim L $ (dashed red and black lines).}
\label{SC}
\end{figure}

To exemplify this we note that the energy of the states at $0$ and $\pi$ only depends on the width of the ribbon, being due to the exponential overlap of the two MBS on the edges, and is given by\cite{Wakabayashi2010} $\epsilon_{0,\pi}\sim\e^{-L_w/\tilde L}$, with $\tilde L$ being the localization length of the MBS and $L_w$ is the width of the ribbon. Hence, $\lim_{L_w=L_\parallel=L\to\infty}\epsilon_{0,\pi}/\lambda=\lim_{L\to\infty}L^2\e^{-L/\tilde L}/4 \pi^2=0$, with $\lambda=4 \pi^2/L^2$ is the mean level spacing. However, for the bands crossing zero away from $k_\parallel=0,\pi$, the lowest energy states typically have a dominant contribution $\epsilon\sim 2 \pi/L_\parallel$, and thus obey $\lim_{L_w=L_\parallel=L\to\infty}\epsilon/\lambda\sim L$. Thus these states never appear as exact-zero energy states. An example for a $\nu=-5$ phase with one MBS and 4 additional zero energy crossings is shown in Fig~\ref{SC}, the same parameters as for Fig.~\ref{BS}(c) are used. The crossing at $k_\parallel=0$, depicted in blue, shows clear exponential scaling to zero relative to the mean level spacing consistent with $\epsilon/\lambda \sim L^2\e^{-L/\tilde L}$. The alternative crossings, depicted in red and black, do not scale to zero, but rather diverge as a power law consistent with $\epsilon/\lambda \sim L$.

\subsection{Gap closing arguments}
Moreover, we can find additional arguments for the allowed values of momentum for which one can form MBS. Thus, considering $k_\parallel$ as a parameter, if there are MBS present for a particular value of $k_\parallel$, one of two things must occur at this particular value of $k_\parallel$. Either the band gap must close or the symmetry of the model must change at these points. In the absence of the first possibility, when the 2D model is fully gapped for a given set of parameters, MBS can only exist at the high symmetry points, which for the current model are $k_\parallel= 0,\pi$.
However situations in which the bulk gap closes as a function of $k_\parallel$ and for which MBS can exist at arbitrary values of $k_\parallel$ may occur. One example is presented in Ref.~\onlinecite{Sedlmayr2015} for a slightly different model, allowing for example for the formation of flat band of MBS in between these special $k_\parallel$ points at which the gap closes. 

The variation of the Chern number with the parameters in our model ($B$ and $\Delta$) was described in Fig.~\ref{numchern}. However, we note that the change in the Chern number, while indicating that the number of edge bands is changing, is not always equivalent to a change in the number of the real MBS states. We argue that a change in the Chern number corresponding to a gap closing at any point which does not correspond to the (1D) TRI momenta cannot change the number of MBS. This includes the Dirac points for the case of a zigzag nanoribbon. In the bandstructures in Fig.~\ref{BS}(c,d) for which we observe 1 MBS on the zigzag edge and 3 on the armchair edge, the 2 additional MBS in the armchair case originate in the gap closing at the Dirac points. Such gap closing  at the Dirac point do correspond to a change in the number of MBS for an armchair ribbon, as in this configuration this corresponds to $k_\parallel=0$. Nevertheless, since the existence of such states depends on the direction of the ribbon, the extra MBS are not fully stable, but rather an example of weak topology.  The $\Gamma$ points in the 2D BZ are TRI momenta for both types of nanoribbon, and therefore a closing of the gap at these points yields a change in the number of MBS. 

\subsection{Majorana polarization arguments}
To test whether the edge states are MBS we can use, along with the energy of the states, the Majorana polarization vector $\hat C (\vec r)=\langle\Psi|\mathcal{C}{\hat r}|\Psi\rangle$.\cite{Sedlmayr2015b,Sticlet2012,Sedlmayr2016,Kaladzhyan2016b,Kaladzhyan2017} As an MBS state is an eigenstate of the particle hole operator $\mathcal{C}$, a Majorana-like state localized inside a spatial region $\mathcal{R}$ must satisfy $C=1$ where $C$ is the normalized magnitude of the integral of the Majorana polarization vector over the spatial region $R$:
\begin{equation}
C=\frac{\left|\sum_{{\vec r}\in \mathcal{R}}\langle\Psi|\mathcal{C}{\hat r}|\Psi\rangle\right|}{\sum_{{\vec r}\in \mathcal{R}}\langle\Psi|{\hat r}|\Psi\rangle}\,.
\end{equation}
Here ${\hat r}$ is the projection onto site ${\vec r}$, and the local Majorana polarization vector $\hat C (\vec r)$ is simply the expectation value of the local particle-hole transformation:
\begin{equation}
\hat C (\vec r)=\langle\Psi|\mathcal{C}{\hat r}|\Psi\rangle=-2\sum_\sigma\sigma u_{{\vec r}\sigma}v_{{\vec r}\sigma}\,.
\label{mp1}
\end{equation}
Here we have written the real space wavefunction in Nambu space as $\Psi_{\vec r}=(u_{{\vec r}\uparrow},u_{{\vec r}\downarrow},u_{{\vec r}\downarrow},v_{{\vec r}\uparrow})$.
We note that in momentum space this mixes eigenstates of different momenta as the conjugation in the particle-hole transformation obeys $\hat{K}\psi(\vec k)=\psi^\dagger(-\vec k)$ where $\psi(\vec k)$ is a wavefunction in momentum space.

In our case where we are interested in nanoribbons we have wavefunctions in a mixture of representations with both spatial and momentum dependence. We then find that
\begin{equation}
\hat C (\vec r)=-2\sum_\sigma\sigma u_{x,k_\parallel,\sigma}v_{x,-k_\parallel,\sigma}\,,
\label{mp2}
\end{equation}
for wavefunctions given by $\Psi_{x,k_\parallel}=(u_{x,k_\parallel,\uparrow},u_{x,k_\parallel,\downarrow},u_{x,k_\parallel,\downarrow},v_{x,k_\parallel,\uparrow})$ with $x$ the position and $k_\parallel$ the momentum.

This is a direct test of whether the states in question are MBS and we will apply this test to our candidate MBS states. Thus, for the examples in Fig.~\ref{BS}, see the Tables \ref{table_mp} and \ref{table_mp2} where we list the energies and the Majorana polarizations for the lowest positive energy states corresponding to various bands. We note that the states at $k_\parallel=0$ have the lowest energies and a $C=1$, indicative of being MBS, and consistent also with previous arguments. 

\begin{table}
\begin{ruledtabular}
{ \renewcommand{\arraystretch}{1.5}
 \renewcommand{\tabcolsep}{0.2cm}
\begin{tabular}{c|c|c|c|c|c|c}
$\nu$ & Edge & $\mu$ & $B$ & $k_\parallel$ & $\epsilon$ & $C$\\\hline
$4$ & ZZ & $0.1t$ & $1.4t$ & $-\frac{4\pi}{5}$ & $6.81\cdot10^{-4}t$  & $0.0725$ \\\hline
$4$ & AC & $0.1t$ & $1.4t$ & $-\frac{29\pi}{110}$ & $1.61\cdot10^{-3}t$  & $0.404$ \\\hline
 $-5$ & ZZ & $1.3t$ & $1.5t$ & 0 & $\Or(10^{-13})t$ & 1\\\hline
 $-5$ & ZZ & $1.3t$ & $1.5t$ & $-\frac{11\pi}{20}$ & $3.84\cdot10^{-4}t$ & $0.613$\\\hline
 $-5$ & AC & $1.3t$ & $1.5t$ & 0 & $\Or(10^{-8})t$ & 1\\\hline
 $-5$ & AC & $1.3t$ & $1.5t$ & $-\frac{29\pi}{200}$ & $1.19\cdot10^{-3}t$ & $0.218$
\end{tabular}}
  \caption{\label{table_mp} The energies  $\epsilon$ and Majorana polarization $C$ for the lowest positive energy states corresponding to the different band crossings in Fig.~\ref{BS}. We take the length of the system to be $L_\parallel=400$ unit cells and its width $L_w=160$. ZZ refers to a zigzag nanoribbon and AC to an armchair nanoribbon.  Note that the two additional crossings for $k_\parallel=0$ in the last armchair case also have an exponentially small energy and $C=1$.}
\end{ruledtabular}
\end{table}
\label{table1}
\begin{table}
\begin{ruledtabular}
{ \renewcommand{\arraystretch}{1.5}
 \renewcommand{\tabcolsep}{0.2cm}
\begin{tabular}{c|c|c|c|c|c|c}
$\nu$ & Edge & $\mu$ & $B$ & $k_\parallel$ & $\epsilon$ & $C$\\\hline
$-2$ & ZZ & $0.5t$ & $1.5t$ & $-\frac{41\pi}{100}$ & $1.03\cdot10^{-4}t$  & $0.392$\\\hline
$-2$ & AC & $0.5t$ & $1.5t$ & $0$ & $\Or(10^{-7})t$  & $1$\\\hline
$-2$ & AC & $0.5t$ & $1.5t$ & $-\frac{11\pi}{60}$ & $1.68\cdot10^{-4}t$  & $0.491$
\end{tabular}}
  \caption{\label{table_mp2} The energies  $\epsilon$ and Majorana polarization $C$ for the lowest positive energy states corresponding to the different band crossings in Fig.~\ref{BS2}. The energies for the additional band crossings at $k_\parallel=0$ for the armchair case are exponentially small, and the corresponding states have $C=1$.}
\end{ruledtabular}
\end{table}

The other states have smaller $C$'s, however this does not automatically imply that they are not MBS since they usually come in pairs, and as the spectrum is degenerate we could also consider combinations of states that can give rise to a maximal $C$.\cite{Sato2006,Sato2009a} One can rule out their MBS character via disorder tests, such as is the case for the non-protected crossings which occur in Fig.~\ref{BS2}(b). However this is not so straightforward for other states. Among the possible combinations we can take we can for example consider $\psi(k_\parallel)$ and $\psi(-k_\parallel)$, however, as these states are localized on different edges in this chiral system it does not increase $C$ (which is indicative of the electron-hole overlap)\cite{Sedlmayr2015b,Sticlet2012,Sedlmayr2016}.
Alternatively one can combine states on the same edge, which belong to different bands, for example the left most and right most green bands in Fig.~\ref{BS}(a). Indeed in this case one obtain a larger Majorana polarization, reaching values of $C\approx0.9$ for this example (with $L_\parallel=400$ and $L_w=160$). However for a finite-size system these states do not have the same energy (here the energy difference between them is $0.00481$t), neither are exact combinations of $k_\parallel$ and $-k_\parallel$, so they are not exact particle-hole eigenstates. Nonetheless in the continuum limit their superposition will become an eigenstate of the particle-hole operator, and their energies become degenerate and equal to zero, so in this limit we cannot distinguish between the real MBS and the non-MBS, and we need to refer back to our scaling arguments to differentiate them.

\subsection{Disorder}

\begin{figure}
\includegraphics[width=0.48\linewidth]{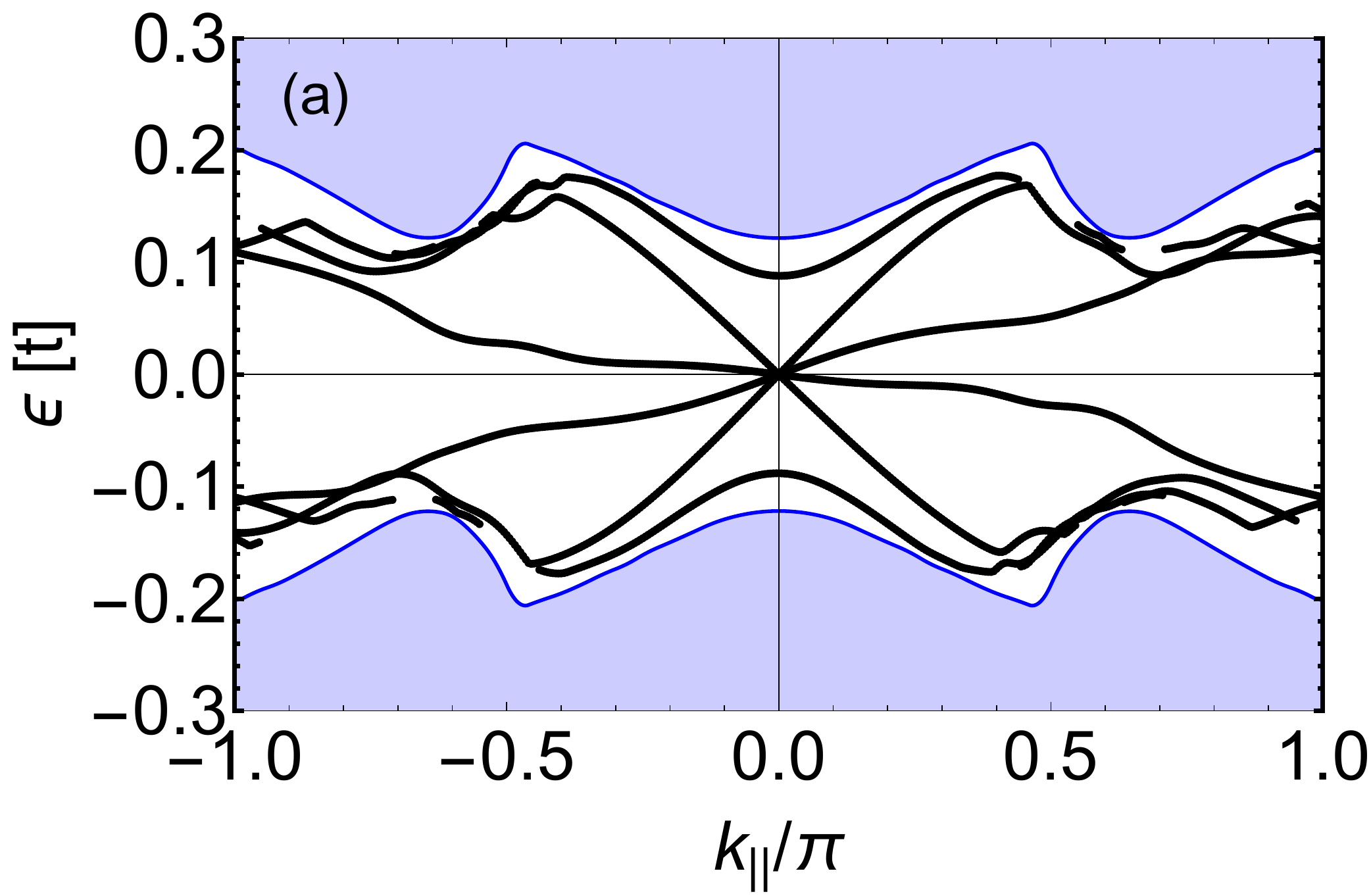}
\includegraphics[width=0.48\linewidth]{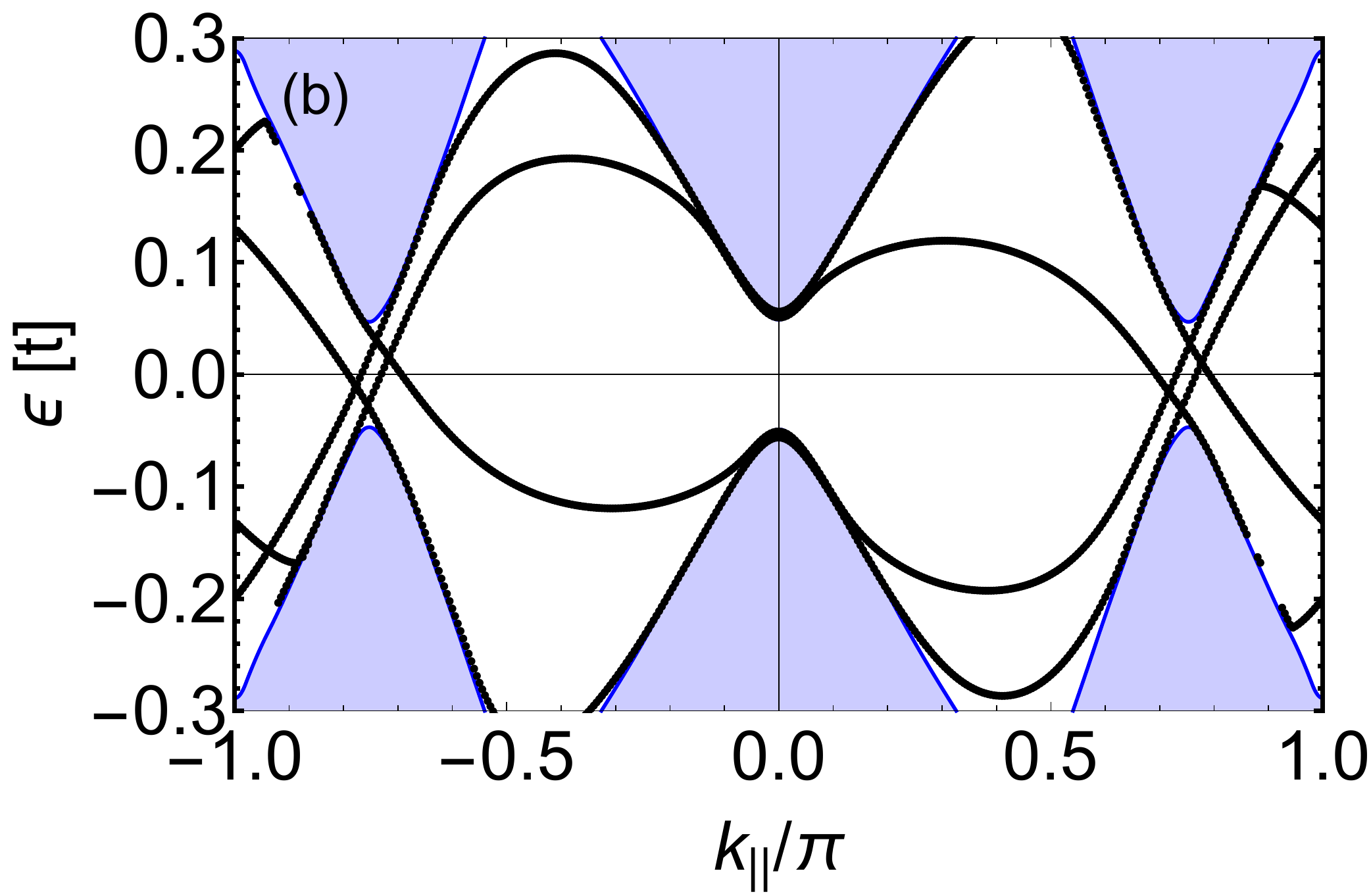}
\caption{(Color online) (a,b) The bandstructures of two armchair nanoribbons: (a) in a regime with $\nu=-2$ and (b) in a regime with $\nu=4$. The parameters are the same as for Fig.~\ref{BS2}(b) and Fig.~\ref{BS}(b) respectively with a disorder strength of $s=0.3$t (a) and $s=0.1$t (b). The bulk bands are those for the clean systems, provided here for reference.}
\label{BS_Dis}
\end{figure}

To further test the nature of the bands crossing between the bulk states we introduce some disorder. We consider an onsite electronic potential which fluctuates randomly along the direction perpendicular to the nanoribbon edge and taking value between $-s\to s$, thus ensuring that $k_\parallel$ remains a good quantum number. In Fig.~\ref{BS_Dis} we present two exemplary cases, with the same parameters as Fig.~\ref{BS}(b) and Fig.~\ref{BS2}(b) for $s=0.1$t. Typically in clean systems the bands of left and right moving electrons cross zero energy and each other simultaneously. In the disordered case this is no longer the case, and indeed for this particular form of disorder, left- and right-movers become mixed, see Fig.~\ref{BS_Dis} and the band are shifted up and down in energy. However, we cannot argue that this corresponds to an obvious destruction of the MBS, the low-energy crossings preserve the same character as in the non-disordered case and the same arguments as above can be applied to justify that the low-energy states are not real MBS.

Nevertheless we can find some zero-energy states that are destroyed by disorder, for example in Fig.~\ref{BS_Dis}(a)  we can see that the non-protected crossings described in Fig.~\ref{BS2}(b) are actually gapped when introducing disorder, confirming the fact that if a band can be continually deformed to eliminate a zero-energy crossing, such crossing is not topologically protected, and thus does not give rise to stable MBS.\cite{Dmytruk2017}

\section{Discussion and conclusions}\label{sec_conclusions}

We have analyzed the relation between the Chern number or equivalently the number of topologically protected edge bands predicted by the bulk-boundary theorem for a 2D topological superconductor, and the number of MBS present along nanoribbon edges. We show that for a lattice model, contrary to expectations, a topologically protected band crossing zero energy does not necessarily contain a state which has full MBS properties. We illustrate this point with several examples for a topological superconductor on a hexagonal lattice. The existence of an MBS inside a given edge band is quite subtle and we provided various arguments that allow one to identify a real MBS from a non-MBS zero-energy crossing. We showed for example that for this model the real MBS appear at high-symmetry points in the band structure, in this case the TRI momenta, while zero-energy crossings occurring at arbitrary momenta can be characterized by scaling arguments to be non-MBS. Thus, the edge states forming close to Dirac points, and reported to be MBS in Ref.~\onlinecite{Wang2016} are not actually real MBS. 

An open question is whether the results presented here also apply to systems which do not allow for a labelling by a transverse ribbon momentum, such as systems containing vortices, or presenting inhomogeneity either in the bulk or at the edges.

While this work was being finalized we became aware of a related work\cite{Liu2017} studying the relationship between Chern numbers and MBS in $p$-wave superconductors.

\acknowledgments Support for this research at Michigan State University (N.S.) was provided by the Institute for Mathematical and Theoretical Physics with funding from the office of the Vice President for Research and Graduate Studies.

\end{document}